\documentclass[11,aip,graphicx,jap,reprint,fixfloat,letterpaper]{revtex4-1}

\usepackage{dcolumn}
\usepackage{bm}
\usepackage{graphicx}

\begin{document}


\title{ NMR Measurements of Power-Law Behavior in the Spin-Wave and Critical Regions of Ferromagnetic EuO}



\author{N. Bykovetz}
\affiliation{Department of the Army,
CECOM LCMC, AMSEL-SF-R,
Fort Monmouth, NJ 07703-5024
}

\author{B. Birang}
\altaffiliation[]{Current address unknown}
\affiliation{Department of Physics, Brandeis University, Waltham, Massachusetts 01742, USA}

\author{J. Klein}
\affiliation{Physics and Astronomy,
University of Pennsylvania,
Philadelphia, PA 19104-6396}

\author{C.L. Lin}
\email[email:]{clin@temple.edu}
\affiliation{Department of Physics,
Temple University,
Philadelphia, PA 19122-6082
}


\date{\today}

\begin{abstract}


Precision continuous-wave NMR measurements have been carried out over the
entire magnetization curve of EuO and are presented in tabular form.  Two very
closely spaced resonances are observed and are attributed to domain and
domain-wall signals.  Both of the signals are useful for analysis in the spin-wave region.
Only the domain signal is measurable above $\sim$50K.  The latter is used for fitting $T_c$ and the
critical exponent $\beta$.  The critical-region fits agree with previous measurements, within
experimental error.  The low-temperature data exhibit a clear-cut $T^2$ behavior, at variance
with the expectations of conventional spin-wave theory.  This result is discussed in relation
to two semi-empirical spin-wave schemes, one formulated by N. Bykovetz, and one by U. Koebler.
The NMR signal at 4.2K gives no indication of a quadrupole splitting, in contradiction to the
interpretation of several previous spin-echo NMR spectra observed in EuO.  This issue remains unresolved.


\end{abstract}


\maketitle 

\vspace{-20 pt}
\section{Introduction}
\vspace{-12pt}

    Precision continuous-wave (CW) nuclear magnetic resonance (NMR) measurements over
the entire magnetization curve of EuO have been carried out and the behavior of these
data is analyzed both in the spin-wave and in the critical regions.  The low-temperature
data exhibit a clear-cut $T^2$ behavior, at variance with the expectations of conventional
spin-wave theory. The critical region power-law fit, however, agrees with other previous
measurements, within experimental error.

    Currently, there is still great interest in the study of EuO because of the varied and
multifaceted characteristics it displays. Among others, oxygen-rich or Gd-doped EuO shows
colossal magnetoresistance, while Eu-rich EuO exhibits a change in conductivity of 13 orders of magnitude
in its insulator to metal transition$^1$. 
Some current studies of EuO have a practical angle (spintronics),
while others focus on testing theories (such as the Kondo-lattice model). While much is not
yet clear in this broader area, the fundamental behavior of the localized spin (pure Heisenberg)
model in EuO is also being put into question by recent measurements$^2$. The results of the current
low-temperature data will hopefully shed additional light on this latter issue.  The principal
focus of this paper is the unconventional  (spin-wave) behavior of
EuO at low temperatures in the context of a pattern of such behavior in other simple magnetic systems.  In addtion, 
the critical-region parameters are examined.

    While a strict $T^2$ low-temperature behavior in pure EuO was reported$^3$ early on, those
measurements were not backed up with numerical evidence. The current measurements are presented in
tabular form (see Table~\ref{tab:datatable}) and show a linear-fit to $T^2$ with an excellent $R^2 = 0.9999$.
In recent years, two alternative semi-empirical spin-wave schemes have been proposed to explain the
simple power-law behaviors displayed by magnetic systems (in many cases not recognized by the original authors),
beginning with the very first NMR measurement of a magnetically ordered system,
$ \text{CuCl}_{2} \cdot$ 2H$_2$O$^4$.  One scheme was developed by U. Koebler$^5$,
and one by N. Bykovetz$^{6, 7, 8}$.  The $T^2$ behavior found in the current EuO NMR data is
consistent with both schemes.

\begingroup
  \squeezetable
  \begin{table}[ht]
    \caption{\label{tab:datatable}NMR Frequency (MHz)\textit{vs}. $T (K)$ for $^{153}$EuO}
    \begin{ruledtabular}
      \begin{tabular}{ccc||cc||cc}

 $T_{H}$&$Freq$&$T_{L}$&$Freq$&$T_{L}$&$Freq$&$T_{L}$\\\hline
4.200&138.500&&69.703&62.519&41.715&67.570\\
13.981&136.080&&68.562&62.807&40.151&67.741\\
17.990&134.190&17.830&66.654&63.286&38.681&67.882\\
21.269&132.240&21.127&65.730&63.507&38.654&67.888\\
25.850&129.050&25.701&63.776&63.973&37.674&67.974\\
26.991&128.090&26.838&60.773&64.608&37.230&68.015\\
28.637&126.715&28.491&58.989&64.976&36.833&68.055\\
32.188&123.523&32.014&57.067&65.347&36.471&68.082\\
34.360&121.370&34.189&53.371&65.993&35.736&68.144\\
38.075&117.352&37.945&52.340&66.160&34.782&68.221\\
38.926&116.350&38.773&51.135&66.357&34.442&68.250\\
42.402&112.020&42.277&51.019&66.369&34.121&68.271\\
43.180&111.000&43.023&49.248&66.643&33.663&68.312\\
47.488&104.720&47.357&47.544&66.852&33.086&68.357\\
&98.822&50.873&46.182&67.044&32.585&68.397\\
&92.680&54.130&46.088&67.060&31.903&68.439\\
&78.507&59.914&44.385&67.266&31.812&68.445\\
&73.502&61.444&43.945&67.322&31.748&68.451\\
&71.795&61.933&43.275&67.399&31.600&68.460\\
&70.723&62.223&43.154&67.411&&\\

      \end{tabular}
    \end{ruledtabular}
  \end{table}
\endgroup

    Using the scheme of Bykovetz, it is argued here that the $T^2$ behavior seen in the EuO NMR
constitutes evidence against a pure localized-spin model as conceived in the conventional Dyson
picture and formalism of magnetic interactions in ferromagnets.

    Our NMR determination of critical parameters agrees with previous results using non-NMR methods$^{9, 10}$.
However, it is an open question why the two so-called ideal Heisenberg ferromagnets EuO and EuS
do not have more nearly identical critical exponents, $ \beta $ ($ \beta = 0.368 $ in EuO and $0.33$ in EuS).
EuS appears to exhibit the same $\beta$ as does MnF$_2$$^{11}$,at variance with most isotropic 3D magnets.

    Lastly, we note that a discrepancy still appears to exist between CW and spin-echo NMR measurements 
in EuO.  The current NMR data, like the original CW data of Boyd$^{12}$, give no indication of 
significant quadrupole splitting of the NMR resonance, whereas spin-echo measurements, which involve 
multi-layered processing of the
raw data$^1$, appear to show the presence of a substantial electric field gradient. An extensive discussion
of this issue was presented in Ref. (1), but the issue still remains unresolved.

\vspace{-12 pt}
\section{Experiment}
\vspace{-12 pt}

NMR measurements on polycrystalline EuO were carried out using a frequency-modulated 
spectrometer in zero external field.
The temperature was determined by means of a platinum resistance thermometer.  
Because of the strong broadening
of the NMR signal on approaching $T_c$, it was decided to carry out the measurements by controlled 
temperature sweeps at fixed frequencies (similar to the method used in references (11) and (13)).  
For consistency, all data, except the helium-bath (4.2K) data point, were determined in this way.

\vspace{-12 pt}
\section{RESULTS AND DISCUSSION}
\vspace{-12 pt}

\begin{figure}[ht]
  \begin{center}
    \includegraphics[width=\columnwidth]{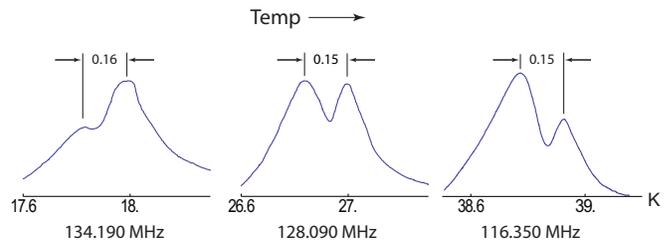}
  \end{center}
  \begin{flushright}
  \caption{\label{fig:scans}NMR temperature-scans of EuO at fixed frequencies at $\sim$17, 27, and 39K, 
shown with the intensities normalized to the same height.  The left peak ($T_L$) 
increases while the  right peak ($T_H$) decreases with increasing temperature.}
  \end{flushright}
  \vspace{-25 pt}
\end{figure}

A single narrow ($\sim$25 kHz) NMR line was observed at 4.2K by sweeping the frequency, as was the case in the
original CW measurement of Boyd$^{12}$.  No indication of quadrupolar splitting was evident.  At temperatures
above 17K, where the frequency was kept constant, and the NMR signal observed by sweeping the temperature,
two closely spaced and overlapping signals were observed all the way up to $\sim$50K.  The signal that peaks
at the lower temperatures will be designated as $T_L$, and at higher temperatures as $T_H$.
Fig. 
\ref{fig:scans} 
shows the behavior of the $T_H$ signal relative to $T_L$ at three representative temperatures.
One can see that the $T_L$ peak tapers off in intensity relative to the $T_H$ peak and becomes undetectable at
4.2K, whereas the $T_H$ peak becomes progressively smaller as 50K is approached, and is undetectable above
this temperature where only the $T_L$ signal is observable.  Near 27K, the signals have equal intensities.

    The $T_H$ signal is ascribed to NMR from domain-walls, since the domain-wall enhancement of NMR
signals is strongest at lowest temperatures.  The  $T_L$ signal must, therefore come from the nuclei in
the domains, since the domain walls become less extensive and their enhancement factor decreases 
as $T_c$ is approached.
Comment, et al.$^1$, presented an argument that in EuO, the domain-wall signal enhancement should be relatively
small,
but our results do not seem to support this contention.

Additional support for the assignment of the two NMR signals as originating in domain and 
domain-wall nuclei comes from the temperature independence of the separation of the peaks in the doublet.
It can be seen from the data in Table \ref{tab:datatable} that the separation $\Delta T = T_H - T_L$
between the observed double peaks (cf., Fig 
\ref{fig:scans}
) is constant, with 
$\Delta T = 0.149 \pm 0.0124$K (on the average), over the temperature range where both signals could be measured.  
As will be seen, both the $T_H$ and the $T_L$ curves give an excellent fit to a $T^2$ power-law.
Consequently, it can be shown that a constant temperature separation translates into a curve separation, 
$ \Delta \nu \propto T$.
A similar $ \Delta \nu = \nu_H - \nu_L \propto T$ separation  was observed in CrBr$_3$,$^{14}$ where the identification 
of which resonance came from domain walls
and which from within the domains was more clear-cut.  It is to be noted that, as a consequence, 
if  one of the curves ($\nu \textrm{~vs.~} T_L \textrm{~or~} T_H$)
goes as $BT^2$,
the other cannot be a pure $T^2$ curve, but must have an additional small corrective term, 
$(2B \Delta T)T$.
It is conjectured that this term should be part of the $T_L$ curve, although the data are not precise enough to
 make a
definitive determination using fitting techniques.

    Focusing on data in the critical region, the following can be seen.  Fits$^{15}$ of the data (to the standard
equation $D(1-T/T_{c})^\beta$, re-cast in the form of a power law) give a $\beta$ of $0.366$  for $T$
data points within $10\%$ of $T_c$, and a $T_c$ of 69.23K.  The parameter $D = 1.18$.  Fitting only points closer
to $T_c$, results in a small, but not a significant increase in $\beta$ as well as $T_c$.  Thus, for the
temperature interval within $3\%$ of $T_c$, $\beta$ becomes $0.378$ and $T_c$ 69.28K.  
The determinations of $\beta$ agree with previous macroscopic$^9$ and neutron-scattering$^{10}$ results, but are 
at odds with the Mossbauer$^{16}$ determination.  
A check for frequency-pulling effects near $T_c$ was made by measuring the temperature 
dependence of the ratio of NMR frequencies in $^{151}$EuO to that in  $^{153}$EuO  (see Table \ref{tab:ratios}). 
The lack of a systematic dependence of the ratio on temperature proves that any such effect is negligible.
The value of the ratio of $^{151}$Eu to $^{153}$Eu frequencies is, however,
lower than in EuS$^{13}$, but agrees with the EuO measurement at 4.2K in Ref. 1 (see footnote 40).

\begingroup
\squeezetable
\begin{table}[ht]
  \caption{\label{tab:ratios} $T$ vs. $\nu$ for $^{153}\text{Eu and } ^{151}\text{Eu}$}
   \begin{ruledtabular}
   \begin{tabular}[t]{cccc}
      $T(K)$&$\nu_{153}(MHz)$&$\nu_{151}(MHz)$&$ {\nu_{151}} / {\nu_{153}}$\\\hline
      67.971&37.718&84.875&2.250\\
      65.811&54.451&122.665&2.252\\
      65.036&58.690&132.312&2.254\\
      64.566&60.991&136.741&2.242\\
  \end{tabular}
   \end{ruledtabular}
\end{table}
\endgroup

\begin{figure}[ht]
\begin{center}
\includegraphics[width=3 in]{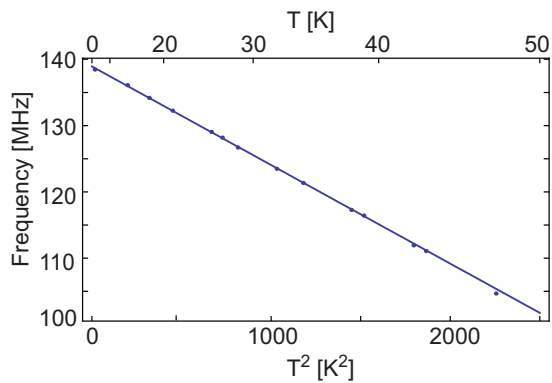}
\end{center}
\begin{flushright}
\caption{\label{fig:T2}$\nu\ vs.\ T^{2}$ in the spin-wave region.  
Deviation below the $T^2$ line begins above $\sim$43K.}
\label{fig:EuOT2}
\end{flushright}
\vspace{-15 pt}
\end{figure}

    In the low-temperature (spin-wave) regime, both the $T_H$ and the $T_L$ data
can be fitted with a simple power-law of the form $\nu = A + BT^c$.  
For the domain-wall, $T_H$, data, one obtains a power law with an exponent of 
$2.04 \pm 0.01$, by including all data points, except the one at 47.488K, 
which is already showing deviation to below the spin-wave $T^2$ curve. 
Fig 
\ref{fig:T2} 
shows a plot of the $\nu$ vs. $T_H^2$, 
displaying the excellent fit.
When this data set is fit to a $T^2$ power law, one gets
$\nu_0 = 138.93$MHz, and an $R^2$ of $0.99992$.
For the $T_L$ data below $\sim43$K, the best-fit exponent is $2.035 \pm 0.011$.  When
fitted to a straight line as a function of $T^2$, one obtains $\nu_0 = 138.96$MHz and an $R^2$ of 0.99994.

    The $T^2$ behavior is in agreement with the expectations of Koebler's scheme$^5$, 
wherein it is claimed that all isotropic 3D magnetic systems should exhibit a $T^2$ 
behavior in the low-temperature region.  
We disagree with the generalization to all magnets, the prime counter-example being the 
$T^{1.71}$ behavior of CrBr$_3$. (see Ref. 6).
However, the $T^2$ spin-wave region behavior of EuO follows the $n=2$ behavior described within the broader
semi-empirical scheme devised by Bykovetz$^6$, wherein both 3D and 2D ferromagnets show power-law behaviors
depicted in Table \ref{tab:exponents}.  
The  power-law behaviors are derivable from the formula
$\Delta M \propto (T^{\frac{3} {2}})^{f(n)}$, where $f(n) = 1/(1-(1/2)^n)$, 
with $n=1,2,3$ for 3D and $n=1,2$ for 2D magnets, where $T^{\frac{3}{2}}$ is replaced by $T$.    
Note that other ferromagnets, e.g., CrI$_3$,$^{17}$  and the near BCC structured compounds, 
$\textrm{M}_2 \textrm{CuX}_{4} \cdot$2H$_2$O$^{18}$, exhibit a $T^2$ behavior,
not noticed by the original authors$^{19}$.

\begingroup
\begingroup
\squeezetable
\begin{table}[ht]
\caption{\label{tab:exponents}Ferromagnetic Exponents}
\begin{ruledtabular}
\begin{tabular}[t]{cccc}
\textbf{n}&\textbf{D=3}&\textbf{D=2}&\textbf{D=1}\\\hline
&~&&\\
3&$T^{1.71}$&&\\
2&$T^{2}$&$T^{1.33}$&\\
1&$T^{3}$&$T^2$&$T$\\

\end{tabular}
\end{ruledtabular}
\end{table}
\endgroup

  The scheme of Bykovetz, if further validated, implies that the conventional Bloch result 
($T^{\frac{3} {2}}$ for 3D and $T^1$ for 2D) determines the power-law behaviors, 
but with an absence of any lattice-dependent effects
(i.e., the Dyson $T^{\frac{5} {2}}$ term, expected in 3D ferromagnets, is missing).  
The general implication would be that even in localized-spin magnetic systems such 
as the europium chalcogenides, the exchange interactions are somehow long-ranged, 
and/or the spin waves propagate in a spin-density-polarized medium (band).  
It is well known that a model of ferromagnetism, in which the magnetization can vary continuously
(as opposed to having discrete spins at localized sites) results in spin waves that display a pure 
$T^{\frac{3} {2}}$ behavior with no ``lattice-correction'' terms.  The recent optical measurements 
by Miyazaki, et al.$^2$, suggest that, in fact, the Heisenberg-Dyson picture of localized exchange 
interactions may not be quite correct.

    Last but not least, the measurements above, 
particularly the 4.2K domain-wall measurement, as well as those of Boyd$^{12}$,
give no evidence of a quadrupole splitting, whereas spin-echo NMR measurements invariably do.
An extended discussion of this problem is given in Ref 1.  
The issue at this point is unresolved.
One could conjecture that the domain-wall resonance in CW measurements somehow averages over 
the electric field gradients.  
On the spin-echo side, the source of the difficulty may lie in the uncertainties of the 
multi-layered analyses of the complex spin-echo
measurements (see Ref 1).


\begin{footnotesize}  
\vspace{.2in}
\noindent
$^1$A. Comment, J. P. Ansermet, C. P. Slichter, H. Rho, C. S. Snow, and S. L. Cooper, Phys. Rev. B \textbf{72
}, 014428 (2005).\\
$^2$H. Miyazaki, T. Ito, H. J. Im, S. Yagi, M. Kato, K. Soda, and S Kimura, Phys. Rev. Lett. \textbf{102}, 22
7203 (2009).\\
$^3$C. Kuznia, H. Pink, and W. Zinn, Colloque Ampere XIV (North Holland Publ. Co., 1967), p. 1216.\\
$^4$N.J. Poulis and G.E.G. Hardeman, Physica \textbf{19}, 391 (1953).\\
$^5$U. K\"{o}bler, A. Hoser, and W. Sch\"{a}fer, Physica B, \textbf{364}, 55 (2005).\\
$^6$N. Bykovetz, J. Klein, and C. L. Lin, J. Appl. Phys. \textbf{105}, 07E103 (2009).\\
$^7$N. Bykovetz, J. Appl. Phys. \textbf{55}, 2062 (1984).\\
$^8$N. Bykovetz, Ph.D. Dissertation, University of Pennsylvania (1976).\\
$^9$B. Menyuk, K. Dwight, and T. B. Reed, Phys. Rev. B \textbf{3}, 1689 (1971).\\
$^{10}$J. Als-Nielsen, O. W. Dietrich, and L. Passell, PR B \textbf{14}, 4908 (1976).\\
$^{11}$P. Heller Phys. Rev. \textbf{146}, 403 (1966).\\
$^{12}$E. L. Boyd, Phys. Rev. \textbf{145}, 174 (1966).\\
$^{13}$P. Heller and G.B. Benedek, Phys. Rev. Lett. \textbf{14}, 71 (1965).\\
$^{14}$A. C. Gossard, V. Jaccarino, J. P. Remeika, J. Appl. Phys. Suppl. \textbf{33} 1187 (1962).\\
$^{15}$All fits in this paper were made using \it{Mathematica}\rm.\\
$^{16}$G. Groll, Zeitschrift fur Physik A \textbf{243}, 60 (1971).\\
$^{17}$A. Narath, Phys. Rev. \textbf{140}, A854 (1965).\\
$^{18}$E. Velu, J. P. Renard, and B. Lecuyer, Phys. Rev. B \textbf{14}(11), 5088 (1976).\\
$^{19}$The low-temperature data from Ref 18 appear in tabular form in Ref 8.

\end{footnotesize}  

\clearpage

\end{document}